\documentclass[aps,prb,reprint,showpacs,amsmath,amssymb,superscriptaddress,floatfix]{revtex4-1}

\usepackage{bm}
\usepackage{bbold}
\usepackage{graphicx}
\usepackage{braket}

\usepackage{color}
\begin{document}

\title{Berezinskii--Kosteriltz--Thouless transition in disordered multi-channel Luttinger liquids}

\author{Max Jones}
\affiliation{University of Birmingham, School of Physics \& Astronomy, B15 2TT, UK}
\author{Igor V. Lerner}
\affiliation {University of Birmingham, School of Physics \& Astronomy, B15 2TT, UK}

\author{Igor V. Yurkevich}
\affiliation{Nonlinearity and Complexity Research Group, School of Engineering \& Applied Science, Aston University, Birmingham B4 7ET, United Kingdom}
\affiliation{Institute of Fundamental and Frontier Sciences, University of Electronic Science and Technology of China, Chengdu 610054, People's Republic of China}

\begin{abstract}
We study the phase transition between conducting and insulating states taking place in disordered multi-channel Luttinger liquids with inter-channel interactions.  We derive renormalisation group equations which are perturbative in disorder but nonperturbative in interaction. In the vicinity of the simultaneous phase transition in all channels, these equations become a set of coupled Berezinskii--Kosterlitz--Thouless equations, which we analyze within two models: an array of identical wires and a two-channel model with distinct channels. We show that a competition between disorder and interaction results in a variety of phases, expected to be observable at intermediate temperatures where the interaction and disorder are relevant but weak hybridization and the charge-density wave interaction may be ignored.
\end{abstract}

\date{\today}


\maketitle

\section{Introduction}

The effect of electron-electron interactions on transport properties of disordered systems {has attracted} a lot of attention since the early prediction \cite{AndersonMIT} and subsequent  renormalisation group (RG) analysis \cite{AALR} of the disorder-driven metal-insulator transition. Weak localization corrections to diffusive transport \cite{AALR,GLKh} are enhanced by the Coulomb interactions signalling further localization of the system \cite{AAL:81}. Interactions in strongly localized systems {lead to a} metal-insulator transition at  {a} finite temperature proportional to the interaction strength (many-body localization \cite{Basko2006}) {suggesting} that the interaction favors delocalization. Experiments on very clean two-dimensional systems show signatures of a metal-insulator transition driven by {a changing} of interaction strength  (for a review, see Ref. \cite{Kravchenko1}).

A theoretical description of the interaction effects in a generic disordered electron system requires non-perturbative approaches{,} which are most developed for one-dimensional systems {whereby} interactions can be treated nonperturbatively  {in terms} of the Luttinger liquid (LL) theory \cite{Giamarchi}. {This} makes it tempting to tackle transport in higher-dimensional anisotropic disordered strongly-correlated systems by making use of the LL model. A promising approach describing rich non-Fermi-liquid physics is to consider an anisotropic system as an array of coupled one-dimensional wires \cite{TK,Po}. Previously,  various exotic states were considered in the framework of the sliding Luttinger liquid (sLL) model,\cite{Sondhi2001,*Vishwanath2001,*MKL2001,*Kane2002} where the RG analysis of the impact of a single impurity embedded into a LL,\cite{KaneFis:92a,*KF:92b,*FurusakiNagaosa:93b} or continuous disorder in LL\cite{GS} has been generalized for a multi-channel case.  A subsequent analysis \cite{SBT} {allowing} for renormalization of the interaction  by disorder has shown that the conducting state does not survive at zero temperature for any realistic long-range inter-wire interactions. The only quantum  phase transition found in Ref.~\onlinecite{SBT} was a superconductor-insulator one with the boundary distorted by  disorder.

A single Luttinger liquid cannot describe {a} metal-insulator transition at high-temperatures where quantum interference does not manifest itself. The only phase transition that is known to happen is the Berezinskii--Kosterlitz--Thouless (BKT) one\cite{Ber1,*Ber2,*KT:73} which takes place when the Luttinger parameter $K=3/2$ (see Ref.~\onlinecite{Giamarchi}) {whereas} repulsive electrons correspond to $K < 1$. The main advantage of {the} sLL model is that inter-wire repulsive interactions stabilize {the} conducting phase{,} bringing {the condition for the BKT transition} into the realm of repulsive fermions. The main disadvantage of the {sLL is that it is unstable (for some non-universal system parameters)} with respect to perturbations like charge-density and Josephson couplings, along with single-particle inter-wire hybridization. These perturbations may become relevant and destroy sLL phase at zero temperature.

In this paper, we focus on the phases existing in the presence of continuous disorder at \emph{finite temperatures} when only the disorder strength  and the electron-electron interaction need to be renormalised, generalizing the recently developed method\cite{IVY:2013,*IVY:2017,*YGYL:2013,*KagLY} based on the scattering matrix formalism.
Since we are not interested in {the} regime of very  low temperatures, where quantum interference governs the transport properties, we may assume that even relevant inter-wire perturbations do not blow up, provided that   their  bare values are sufficiently small  and  temperature  infra-red cutoff is relatively high. This is the case of an array of wires that are well separated from each other (weak hybridization means weak single- and two-particle Josephson couplings) and {the interaction potential between the wires is smooth of the scale of the Fermi wave-length.}, so that the   bare value of inter-wire charge-density-wave interaction is small.

\section{The multichannel model}

The action describing a multichannel LL is a straightforward generalization of the standard LL action:\cite{Giamarchi}
\begin{align}
S&=\frac{1}{8\pi}\,\int{\rm d}x\,{\rm d}t\,{\bm \Psi}^{\rm T}\left[{\hat\tau}_1\,\partial_t +{\mathsf{V}}\,\partial_x\right]\,\partial_x{\bm\Psi}\notag\\[-6pt]\label{S}\\[-6pt]
&+iD\sum_i\int{\rm d}x\,{\rm d}t\,{\rm d}t'\,\cos \left[\theta_i(x,t)-\theta_i(x,t')\right]\,.\nonumber
\end{align}
Here the  composite vector field ${\bm\Psi}^{\rm T}=({\bm\theta}^{\rm T}\,,{\bm\phi}^{\rm T})$
is built on two vector fields, ${\bm\theta}=(\theta_1\,,...\,,\theta_N)$ and ${\bm\phi}=(\phi_1\,,...\,,\phi_N)$, that parametrise density and current excitations in the $i^{{\mathrm{th}}} $ channel \((1\leq i\leq N)\) as \(\rho_i=\frac{1}{2\pi}\partial_x\theta_i\) and \(j_i=\frac{1}{2\pi}\partial_x\phi_i\);
${\hat\tau}_1$ is the Pauli matrix in $\{{\bm\theta}\,,{\bm\phi}\}$-space and ${\mathsf{V}}={\rm diag}[{\mathsf{V}}_+\,,{\mathsf{V}}_-]$ is a block-diagonal (in the same space) matrix describing density-density, ${\mathsf{V}}_+$, and current-current, ${\mathsf{V}}_-$, interactions. In the absence of inter-channel interactions these matrices would become diagonal, $\left[V_{\pm}\right]_{ij}\to\delta_{ij}\,v_i\,K_i^{\mp 1}$, with $v_i$  and $K_i$ being the velocities and  Luttinger parameters in the $i^{{\mathrm{th}}} $ channel.

The nonlinear, cosine term in the action  results from the standard replica averaging over disorder, as in the single-channel case \cite{Giamarchi}, albeit the replica indices are suppressed in Eq.~\eqref{S}. The averaging has been performed over   the standard single-particle disorder potential with random backscattering amplitudes, $\xi_i(x)\,e^{i\theta_i(x)}+{\rm c.c.}$, with the white-noise correlations,
\begin{align}\label{corr}
  \left<\xi_i(x)\,\overline{\xi}_j(x')\right>=\delta_{ij}\,D_ {i}\,\delta (x-x') ,
\end{align}
assuming the absence of inter-channel correlations.

\subsection{RG equations}

Following the standard procedure \cite{Giamarchi}, one derives the following  {RG} equations for {the} disorder strength and interaction matrices (with $l$ being the logarithm of the scaling factor):
\begin{align}\label{RG}
\begin{aligned}
&\partial_{l} {\mathsf{D}}=(3-2\widetilde{\mathsf {K}})\,{\mathsf{D}}\,,\quad&& {\mathsf{D}}(l\!=\!0)=D_0\, {\mathbb{1}}\,;\\
&\partial_{l} {\mathsf{V}}_{-}^{-1}={\mathsf{D}}\,, && {\mathsf{V}}_{-}(l\!=\!0)={\mathsf{V}}_{-}^{(0)}\,.
\end{aligned}
\end{align}

The density-density interaction matrix, $V_+$, does not {renormalize}: $\partial_{l}{\mathsf{V}}_{+}=0$. In  Eq.~\eqref{RG}, we have introduced  matrices ${\mathsf{D}}$ and $\widetilde{{\mathsf{K}}}$ which are diagonal in {the channel space}: ${\mathsf{D}}={\rm diag}\{{ D_1\,,\,...\,,D_N}\} $ and $\widetilde{\mathsf {K}}={\rm diag} \{{K_{11}\,,\,...\,, K_{NN}}\} $. The elements of ${\mathsf{D}}$ describe the disorder strength in the appropriate channel, Eq.~\eqref{corr}, while the elements of $\widetilde{{\mathsf{K}}}$  are the diagonal elements of the \emph{Luttinger matrix},  ${\mathsf{K}}$,  which is defined  from the equation
\begin{equation}\label{K}
{\mathsf{K}}\,{\mathsf{V}}_{+}\,{\mathsf{K}}={\mathsf{V}}_{-}\,.
\end{equation}
 In the $N$-channel RG equations, the Luttinger matrix plays the role  similar to that of the Luttinger parameter in the single-channel LL  (see Ref.~\onlinecite{IVY:2013} for details).
Equation \eqref{K} closes the set of RG equations \eqref{RG}. Below we build the phase diagrams corresponding to these equations for two particular cases.

\section{Lattice of Identical Channels}

 {Here we consider the multi-channel model where identical channels (wires) are packed into a 2D or 3D array in such a way that the cross-section perpendicular to the length of the wires forms a lattice ${\cal L}$}. All matrix elements of {the interaction matrices,} ${\mathsf{V}}_{\pm}${,} may be labelled by the spatial positions ${\bm R}$ of wires in the perpendicular plane, ${\mathsf{V}}_{\pm}\to V_{\pm}({\bm R},{\bm R}')$, where ${\bm R}\subset {\cal L}$. If the lattice is a Bravais one, the matrix elements of ${\mathsf{V}}_{\pm}$  become scalars, $V_{\pm}({\bm R}-{\bm R}')$, (assuming translation invariance and periodic boundary conditions). {For non-Bravais lattices they will become matrices in the space of inequivalent wires, which we do not consider here}. Equation \eqref{K} for the the Luttinger matrix transforms to
\begin{equation}\label{F}
\sum_{{\bm R}_1,{\bm R}_2\subset{\cal L}}\,K({\bm R}_{12})\,V_{+}({\bm R}_{23})\,K({\bm R}_{34})=V_{-}({\bm R}_{14})\,,
\end{equation}
where ${\bm R}_{ij}={\bm R}_i-{\bm R}_j$. This equation is easily solved with the use of the discrete Fourier transform  on the lattice ${\cal L}$,
\begin{align*}
F({\bm R})&=\int \frac{{\rm d}^dq}{(2\pi)^d}\,F({\bm q})\,e^{i{\bm q}{\bm R}}\,,\\
F({\bm q})&=\sum_{{\bm R}\subset{\cal L}}\,F({\bm R})\,e^{-i{\bm q}{\bm R}}\,.
\end{align*}
Here and elsewhere in this section the  momentum integration is performed over the Brillouin zone of the wire lattice. The solution to the Eq.~\eqref{F} has the form
\begin{eqnarray}\label{Kr}
K({\bm r})=\int\frac{{\rm d}^dq}{(2\pi)^d}\,\sqrt{\frac{V_{-}({\bm q})}{V_{+}({\bm q})}}\,e^{i{\bm q}{\bm r}}\,,
\end{eqnarray}
with ${\bm r}={\bm R}-{\bm R}'\subset{\cal L}$ and $V_{\pm}({\bm q})$ being the discrete Fourier transform of interaction potentials $V_{\pm}({\bm r})$.

Now the diagonal matrix $\widetilde{{\mathsf{K}}}$ in Eq.~\eqref{RG} is reduced to $\widetilde{K}{\mathbb{1}}$ where the effective Luttinger parameter ${\widetilde{K}}$
is given by $\widetilde{K}\equiv K({\bm r=\bm0})$.
Assuming equal strength of bare disorder  in each wire, $D_i=D$, the matrix RG equations \eqref{RG} are reduced to the  two RG equations for the disorder strength and the deviation, $c$,  of the Fourier transform of the current-current interaction from its bare value
\begin{align}\label{RGq}
\begin{aligned}
&\partial_{l} {{D}}=\left[3-2\widetilde{ {K}}(c)\right]\,{{D}}\,,\quad&&  \partial_{l} c={{D}}\,;\\
&{{D}}(l\!=\!0)=D_0\,, && c({l\!=\!0})=0\,,
\end{aligned}
\end{align}
with $c({l})$ defined by
\begin{equation}\label{c}
 c(l)\equiv V^{-1}_-({\bm q};l)-V^{-1}_0({{\bm{q}}})\,,
\end{equation}
where $V_0({{\bm{q}}})\equiv V_-({\bm q},l\!=\!0)$ is the bare value of the current-current interaction.
The closure to the RG equations is provided by the explicit dependence of $\widetilde{K}\equiv \widetilde{K}({\bm {r}={\bm{0}}})$ on $c$:
\begin{equation}\label{k}
\widetilde{K}(c)=\int\frac{{\rm d}^dq}{(2\pi)^d}\,V_{+}^{-1/2}({\bm q})\,\left[V_0({{\bm{q}}})+c({l})\right]^{-1/2}\!.
\end{equation}
The effective Luttinger parameter $\widetilde{K}(c)$ is  {a} monotonically decreasing function of $c$ and, therefore, if its bare value $\widetilde{K}_0\equiv \widetilde{K}(c\!=\!0)<3/2$, the disorder will always {grow} under renormalisation and we always end up in the insulating regime. From now on, we will be interested only in  {the} case  $\widetilde{K}_0>3/2$.

The BKT transition takes place at $c=c^{*}$ with the critical value $c^*$ found from $
\widetilde{K}(c^{*})=\frac{3}{2}\,.
$
The analysis of the RG flow in these terms  is possible only in the vicinity of the BKT critical value $\widetilde{K}=3/2$.  {This} means that the bare value $\widetilde{K}_0$  should also be close to $3/2$. In this case $c^{*}\ll 1$ and we may approximate
\begin{equation}
\widetilde{K}(c)\approx \widetilde{K}_0-\kappa\,c\,,
\end{equation}
where
\begin{equation}\label{kappa0}
\kappa=-\left.\frac{{\rm d}\widetilde{K}(c)}{{\rm d}c}\right|_{c\!=\!0}=\frac{1}{2}\,\int\frac{{\rm d}^dq}{(2\pi)^d}\,\frac{V_{0}^{3/2}(\bm q)}{V_{+}^{1/2}({\bm q})}\,.
\end{equation}
The critical value $c^{*}$ is given in terms of the initial detuning from the transition:
\begin{equation}
c^{*}=\frac{\delta}{\kappa}\ll 1\,,\quad \delta\equiv \widetilde{K}_0-\frac{3}{2}\geq 0\,.
\end{equation}
The BKT RG equations acquire the standard form,
\begin{align}\label{x}
\partial_l D&=2x\,D\,,  &
\partial_l x&=\kappa\,D\, , & x&\equiv \kappa\,c-\delta
\end{align}
with the initial condition $ x(l\!=\!0)=-\delta$ in terms  of $x$.
The RG flows in the $(x\,,D)$-plane obey the equation
\begin{equation}
\kappa\, \frac{{\rm d}D}{{\rm d}x}=2x
\end{equation}
that defines the family of trajectories
$
D= {x^2}{\kappa}^{-1} +E\,,
$
with the constant $E$  being defined by the initial values, $
E=D_0- {\delta^2}{\kappa}^{-1}
$.

\begin{figure}
	\begin{center}
		\includegraphics[width=0.45\textwidth]{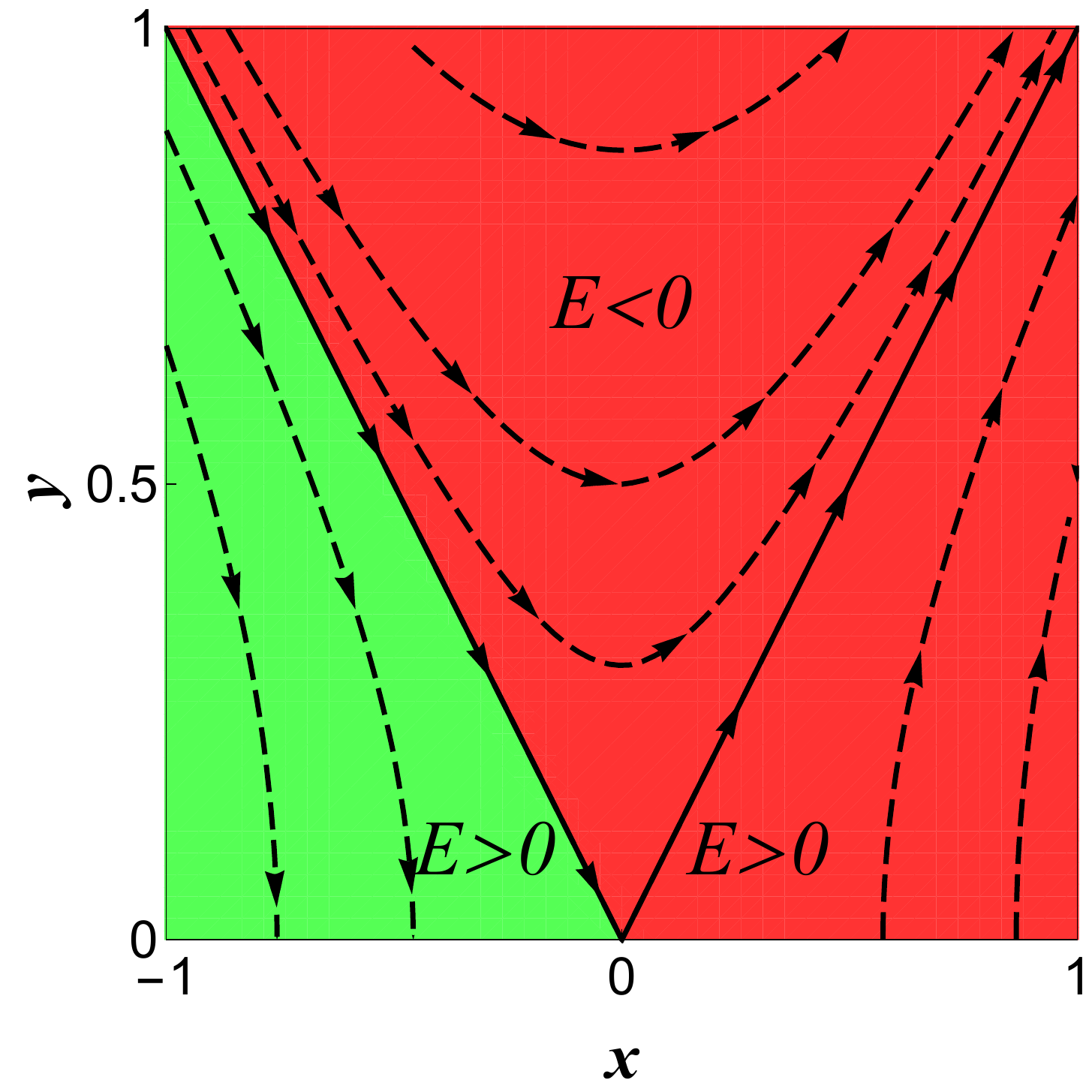}
		\caption{(color online) {BKT flow diagram for the wire lattice, with \(y=\sqrt{\kappa D}\) characterizing the disorder, and \(x\) is the deviation from the critical value of Luttinger parameter \(\widetilde{K}_0=\frac{3}{2}\). The equations have a conserved `energy' \(E\). If \(E<0\) or $E>0$ and $x>0$, the system flows to strong disorder, while only if \(E>0\) and \(x<0\) does it flow to a conducting state.}}
		\label{fig:KTflow}
	\end{center}
\end{figure}

The boundary between the insulating and conducting phases corresponds $E=0$, i.e.\ $
\delta =\sqrt{\kappa\,D_0}\equiv y$:
the system is conducting for $\delta>\sqrt{\kappa\,D_0}$ and insulating for $\delta<\sqrt{\kappa\,D_0}$.
 These RG flows  are illustrated in Fig.~\ref{fig:KTflow}, where the phase boundary is clearly seen.  The effects of inter-wire interactions are in the definitions of the parameters, given explicitly in  Eqs.~\eqref{k}--\eqref{kappa0}.

 The position of the phase-separation boundary is  mainly dictated by the interaction, $\widetilde{K}_0=\frac{3}{2}$, which also governs renormalisation of the (weak) disorder strength while the feedback from disorder to interaction is negligible. One can show\cite{CKY}  that the inter-wire long-range interaction results in $\widetilde{K}_0> K$ (where $K$ is the standard single-wire Luttinger parameter), so that it favors  {a} conducting state. The effective Luttinger parameter $\widetilde{K}_0$ depends on both $K$ and inter-wire interaction parameters and can reach  {the} value $3/2$ even for $K<1$ corresponding to repulsive fermions. Therefore, one should expect a competition between the weak inter-wire long-range interaction and weak disorder leading to the BKT metal-insulator transition for the wire lattice.

\begin{figure*}
\begin{center}	
		{\includegraphics[width=0.45\textwidth]{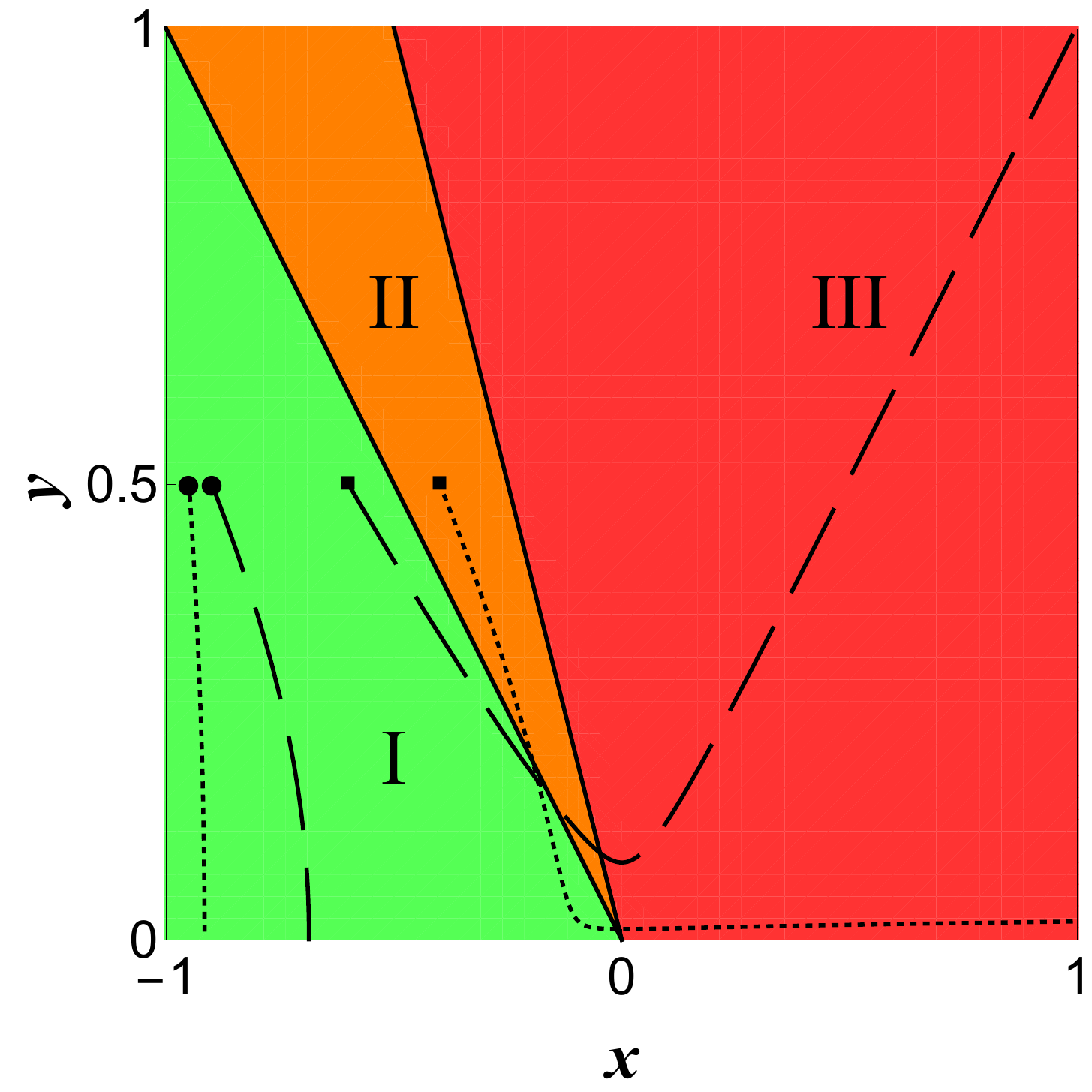}}\qquad\quad
{\includegraphics[width=0.45\textwidth]{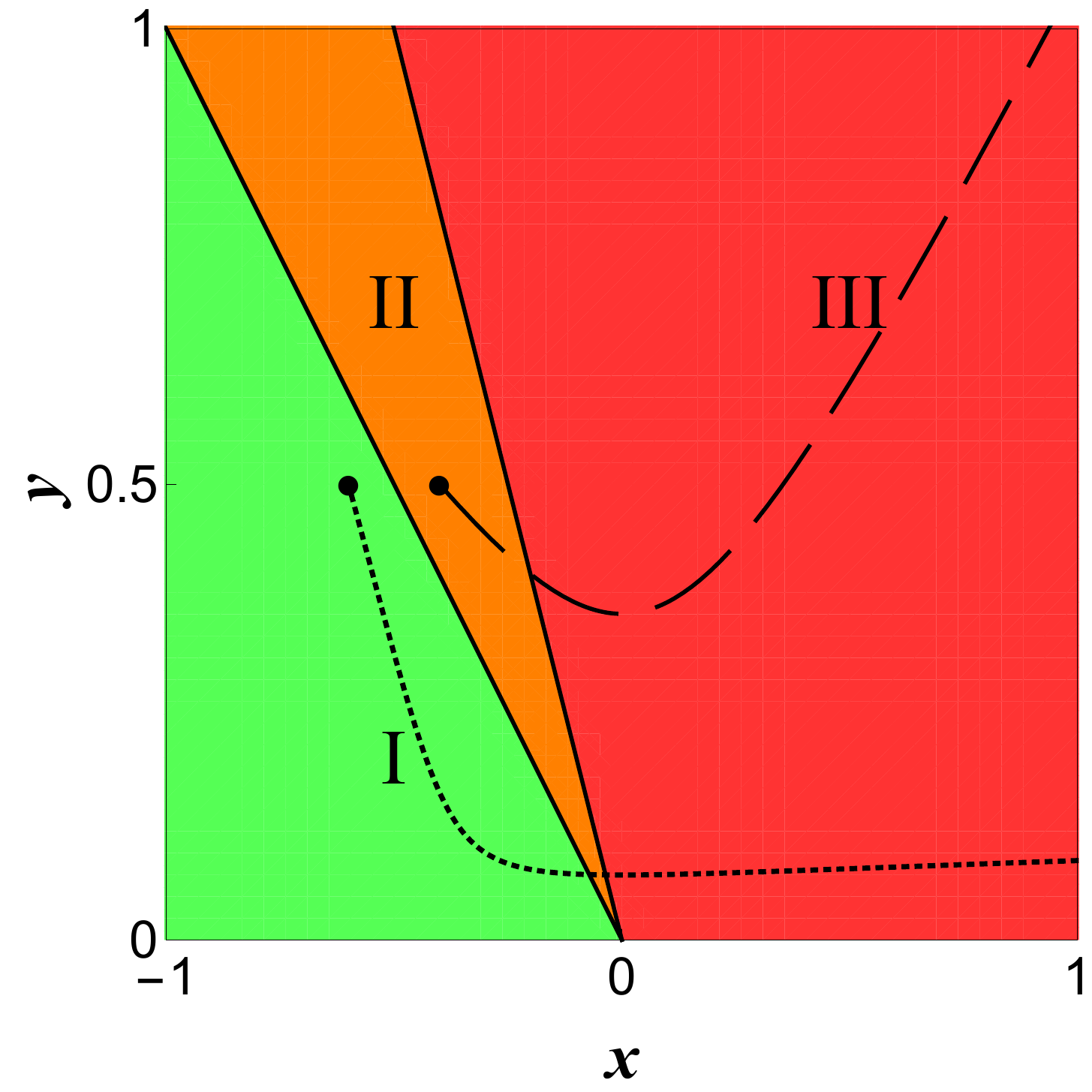}}

\vspace*{5pt}

 \hspace*{11mm}{\large (a)} \hspace*{85mm}{\large (b)}

		\caption{\label{fig:CC}{Phase diagram of a two-channel disordered Luttinger liquid. Differently shaded regions I (green online),  II (amber) and III (red) correspond to three phases of the uncoupled channels:   $cc$  (I),  $ic$ ({II}) and $ii$ ({III}). Three pairs of trajectories that show the effect of  the inter-channel coupling, $\kappa$,  were calculated for \(\kappa =0.4\kappa_{11},\kappa_{22}=0.25\kappa_{11}\) in Eq.~\eqref{kappa1}, and  \(y_0=\left(4\kappa_{11}\right)^{-1/2}\) in Eq.~\eqref{Kondo}, with \(\kappa_{11}\) defining the scale. In each pair, the dashed and dotted lines correspond to channels 1 and 2 respectively. The pair started with circles in (a) corresponds to a system with energy \(E\approx2.1\kappa_{11}\) deep inside the (cc) region where the coupling does not lead to qualitative changes. The pair started with squares in (a) corresponds to \(E\approx 0.07\kappa_{11}\) which is ostensibly in the (cc) region for an uncoupled system;  the coupling generates a large enough negative energy shift to push the system into the (ii) region. The pair in (b) is for \(E\approx 0.91\kappa_{11}\) when the dashed trajectory is in the insulating region of channel 1. As the disorder there  begins to grow,  the dotted trajectory for channel 2 that remains in the conducting region for $\kappa=0$ is dragged into insulating region with it. This shows how the mixed (ic) phase is destroyed by inter-channel coupling.}}
	\end{center}
\end{figure*}

\section{Two distinct channels}
We limit our analysis of non-identical channels with different (and uncorrelated) disorder strength to the case of two channels. Then all matrices in the RG equations \eqref{RG} are  $2\times 2$.
Similar to the approach used in the previous section for the $N$-channel problem, we   {now} introduce two renormalizable scalars, $c_{1,2} $, describing the deviation of the current-current interaction matrix from its bare value, ${\mathsf{V}}_0\equiv {\mathsf{V}}_{-}({l\!=\!0})$:
\begin{equation}
{\mathsf{V}}^{-1}_-(l)={\mathsf{V}}_0^{-1}+\mathsf{c}\,,\qquad \mathsf{c}=\left(
\begin{array}{cc}
c_1 & 0 \\
0 & c_2 \\
\end{array}
\right)\,.
\end{equation}
The RG equations \eqref{RG} in the new  {variables} become
\begin{align}\label{RGc}
\begin{aligned}
&\partial_{l} {\mathsf{D}}=\left[3-2\widetilde{\mathsf {K}}({\mathsf{c}})\right]\,{\mathsf{D}}\,,\quad&& {\mathsf{D}}(l\!=\!0)=D_0\, {\mathbb{1}}\,;\\
&\partial_{l} \mathsf{c}={\mathsf{D}}\,, && \mathsf{c}(l\!=\!0)=0.
\end{aligned}
\end{align}
Again, ${\widetilde{\mathsf{K}}}={\mathrm{diag}}\{{K_{11}, K_{22}  }\} $, with the
 two effective Luttinger parameters $K_{ii}(c_1,\,c_2)$ being the diagonal elements of the Luttinger matrix ${\mathsf{K}}$, Eq.~\eqref{K}. This equation   can now be rewritten via ${\mathsf{V_0}}$ and ${\mathsf{c}}$ as
\begin{equation}\label{KVK}
\mathsf{K}\,\mathsf{V}_+\,\mathsf{K}\,\left[\mathsf{V}_0^{-1}+\mathsf{c}\right]= {\mathbb{1}}
\end{equation}

Assuming   the system to be initially in the vicinity of a generalized BKT transition in each channel, i.e.\   for both $i=1,2$ one has $|K_{ii}^{({0})} -3/2|\ll 1$, one can see that the critical values at which the BKT  transition occurs,
\begin{equation}
K_{ii}\left(c^*_1,c^*_2\right)=\frac{3}{2}\,,\qquad i=1,2\,,
\end{equation}
are small, $|{ c}^{*}_i|\ll 1$.
Therefore, in the vicinity of the transition one may use the expansion
\begin{align}\label{kappa}
K_{ii}(c_1,c_2)&\approx K_{ii}^{(0)}-\sum_j\!\kappa_{ij}\, c_j\,,\;
\quad \kappa_{ij}\equiv -\left.\frac{\partial K_{ii}}{\partial c_j}\right|_{c_1=c_2=0}.
\end{align}
The matrix of derivatives, $\{\mathsf{\kappa}_{ij} \}$, is a symmetric positively definite matrix with the positive matrix elements,
\begin{align}\label{kappa1}
\begin{aligned}
\kappa_{ii}&=T^{-1}\,\left[(1-k^2)\,V_{ii}^2+{\rm det V}\right]\geq 0\,,\\
  \kappa&=T^{-1}\,\left[(1-k^2)\,V_{12}^2+k^2\,{\rm det V}\right]\geq 0\,,
\end{aligned}
\end{align}
 where $\kappa\equiv \kappa_{12}=\kappa_{21}$ (see Appendix A for the derivation).

 The RG equations \eqref{RGc} turn into
\begin{eqnarray}\label{RGxs}
\begin{aligned}
&\partial_{l} D_i=2x_i\,D_i\,,\qquad && \partial_{l} x_i= \sum_{j=1,2}\,\kappa_{ij}\,D_j\,;\\
&D_i(l\!=\!0)=D_0\,, && x_i(l\!=\!0)=-\delta_i\,,
\end{aligned}
\end{eqnarray}
where we have introduced the  notations
\begin{equation}
x_i=\sum_j\kappa_{ij}\,c_j-\delta_i\,,\qquad \delta_i=K_{ii}^{(0)}-\frac{3}{2}\,,
\end{equation}
to present them in the form familiar from the previous chapter. The substitution $D_i=y_i^2$  reduce these  Eq.~\eqref{RGxs}  to the pair of coupled BKT equations:
\begin{align}\label{Kondo}
\begin{aligned}
&\partial_{l} y_i = x_i\,y_i\,, &&
\partial_{l} x_i  = \sum_j \kappa_{ij}\,y^2_j\, ,\\
&D_i(l\!=\!0)=D_0\,, && x_i(l\!=\!0)=-\delta_i\,,
\end{aligned}
\end{align}
 There is only   one integral of motion (see Appendix B for details), given in terms of  ${\bm x}{=}(x_1,x_2)$ and ${\bm y}{=}(y_1,y_2)$ by
\begin{equation}\label{E0}
E={\bm x}\,{{\mathsf{m}}}\,{\bm x}-{\bm y}^2\,,\quad{\mathsf{m}}={\hat \kappa}^{-1}=\left(
                                                                             \begin{array}{cc}
                                                                               m_1 & -m \\
                                                                               -m & m_2 \\
                                                                             \end{array}
                                                                           \right)
\,,
\end{equation}
where  $m_{1,2}$ and $m$ are positive.

In the absence of the inter-channel coupling, $m=0$, Eqs.~\eqref{Kondo} would describe two independent systems undergoing the BKT transition. They are equivalent to two uncoupled Kondo impurities, each having the integral of motion (`energy'),
\begin{equation}\label{E}
E_i=m_i\,x_i^2-y_i^2\,,
\end{equation}
with the exchange constants $J_i^{\perp}\equiv y_i$ and $J_i^{\parallel}\equiv x_i$. { Then $\delta_i<0$ in Eq.~\eqref{RGxs} corresponds to the anti-ferromagnetic Kondo impurity ({$J_i^{\parallel}>0$}) {where} all the RG  flows go towards   the strong-coupling fixed point (the unitary limit of the Kondo screening when $J_i^{\perp}\to\infty $ at the Kondo temperature) corresponding to the insulator. The case of $\delta_i>0$ corresponds to the ferromagnetic Kondo impurity, $J_i^{\parallel}<0$, where the flows go the strong-coupling fixed point for $E_i<0$, but to the weak-coupling fixed point corresponding to the conductor for $E_i>0$. For completeness, these well-known results including explicit expressions for the RG flows  are recouped in Appendix~B.}

Let us analyze an impact of a weak inter-channel coupling, $m\ll 1$, when Eqs.~(\ref{Kondo}) describe two coupled Kondo impurities. In this case, there is only one integral of motion,
Eq.~\eqref{E0},
while the former integrals of motions, Eq.~\eqref{E},  become `adiabatic invariants', i.e.\  slow functions of `time' $l$:
\begin{equation}\label{dotE}
\dot{E}_i=2m\,x_i\,\dot{x}_{-i}\,.
\end{equation}
{In Appendix C, we show how to use these invariants to construct the RG flows starting from the uncoupled case. Here we summarize the results of this consideration.} 

If both $ \delta_i \leq  0$ in Eq.~\eqref{RGxs}, the system flows to the strong-coupling Kondo regime, i.e.\ an insulator. It means that  {the} $(ii)$ phase where both channels were insulators for $m=0$ is not qualitatively affected by the coupling between channels but just expanding in the phase space.

If both $ {\delta_i}\geq 0$, the RG flows  depend on the bare values of the adiabatic invariants $E_i(0)$. They start with a negative derivative, Eq.~\eqref{dotE},  bending  upwards in comparison to those without the coupling. The flows are illustrated in Fig.~\ref{fig:CC} where the trajectories are numerically calculated for several values of the parameters.  Note that the critical value for the initial values of $E_i$  to stay on a conducting trajectory (leading to $y_i(l=\infty)=0$)  increases so that the $(cc)$ phase shrinks.

{The mixed $(ci)$ phase turns out to be totally unstable  as illustrated in Fig.~\ref{fig:CC}. Since the RG trajectories in the insulating channel flow towards the strong-coupling Kondo regime, the negative `energy' shift  in the initially conducting channel arising from the coupling will be sufficient to drag the channel into the negative energy region (See  Appendix  C for details), finally making it insulating as well. Thus the intermediate $ci$ phase eventually disappears due to the inter-channel coupling, while the BKT transition between the (ii) and (cc) phases is shifted towards the insulator. }

\section{Conclusion}
We have constructed a generic description of a `high'-temperature Berezinskii-Kosterlitz-Thouless transition in a multi-channel array of coupled Luttinger liquids. We have focused on the two cases, a lattice of identical LL wires or two distinct LL channels.  The inter-channel coupling makes these transitions in principle observable not only in systems with locally repulsive bosons but also in  systems with repulsive fermions, where no such a transition exists for a single LL channel.

\section*{Acknowledgement} IVY gratefully acknowledges support from  the Leverhulme Trust via Grant No.\ RPG-2016-044 and hospitality extended to him at  the Center for Theoretical Physics of Complex Systems, Daejeon, South Korea.

\appendix
\section{Calculation of  ${\kappa}_{ij} $, the matrix of derivatives}
Instead of differentiating  {the} explicit solution with respect to ${\bm c}$ at the BKT transition, it  {is} easier to start from   linearizing  Eq.~\eqref{KVK} in ${\mathsf{c}}$, using
$
{\mathsf{K}}({\mathsf{c}})={\mathsf{K}}+\delta{\mathsf{K}} \,,
$
where the diagonal elements of ${\mathsf{K}}\equiv {\mathsf{K}}(l\!=\!0)$ are taken at  the BKT transition $K_{ii}=3/2$:
\begin{equation}
{\mathsf{K}}=\tfrac{3}{2}\left(
                      \begin{array}{cc}
                        1 & k \\
                        k & 1 \\
                      \end{array}
                    \right)\,.
\end{equation}
Here    the only restriction on the off-diagonal element of ${\mathsf{K}}$  is $|k|\leq 1$.   Using Eq.~\eqref{KVK} at the BKT point allows us to eliminate  ${\mathsf{V}}_+$:
\begin{equation*}
{\mathsf{V}}_+ =  {\mathsf{K}}^{-1} {\mathsf{V}}_ {0}{\mathsf{K}}^{-1} \,.
\end{equation*}
  Then the linearization results in
\begin{equation*}
{\mathsf{V}}_0^{-1}\,\delta{\mathsf{K}}\,{\mathsf{K}}^{-1}+{\rm transposition}=-{\mathsf{c}}\,.
\end{equation*}
This equation has a solution parametrized by an arbitrary scalar $\omega$,
\begin{equation}\label{results}
\delta{\mathsf{K}}=-\frac{1}{2}{\mathsf{V}}_0 \left[{\mathsf{c}}+\omega\,{\hat\sigma}_2\right]\,{\mathsf{K}}\,,
\end{equation}
where ${\hat\sigma}_2=i\hat{\tau}_2$, and $\hat{\tau_2}$  is a Pauli matrix. The scalar $\omega$ is found from the fact that both ${\mathsf{K}}$ and  $\delta{\mathsf{K}}$
are symmetric matrices.
\newcommand\tr{\operatorname{tr}}
This condition can be written as $\tr {\hat\sigma}_2\delta{\mathsf{K}}=0$ which  results in
\begin{align}
\begin{aligned}
\omega&=-T^{-1} \tr{\mathsf{K}}{\hat\sigma}_2{\mathsf{V}}_0\,{\mathsf{c}}\,,\\
T&=-\tr {\mathsf{K}}{\hat\sigma}_2{\mathsf{V}}_0{\hat\sigma}_2=V_{11}+V_{22}-2k\,V_{12}\geq 0\,,
\end{aligned}
\end{align}
with $V_{ij}$ being the matrix elements of ${\mathsf{V}}_0$.
Substituting this into Eq.~\eqref{results} and differentiating   $\delta{\mathsf{K}}$ with respect to $c_i$ gives  (up to a factor) the matrix ${\kappa}_{ij} $, Eq.~\eqref{kappa}:
\begin{equation}
\kappa_{ij}=K_{ij}\,V_{ji}-T^{-1}\,\left({\mathsf{K}}{\hat\sigma}_2{\mathsf{V}}\right)_{ii} \,\left({\mathsf{K}}{\hat\sigma}_2{\mathsf{V}}\right)_{jj}.
\end{equation}
The  matrix ${\hat\kappa}$ is symmetric with positive matrix elements.
One can easily check that $\det\hat\kappa\geq 0$ so  that both ${\hat\kappa}$ and ${\hat\kappa}^{-1}$ are positive-definite matrices.

\section{Integrals of motion and RG flows}
Equation~\eqref{Kondo}, defines the RG flows in four-dimensional space $\{y_i,x_i\}$ with $i=1,2$.
Since we are interested in the asymptotic behavior of  $D_i\equiv y_i^2$ only, we   eliminate $x_i$  using the first of Eqs.~\eqref{Kondo} to obtain  $x_i=\partial _l \ln y_i$. Introducing the `force' $F_i\equiv D_i=y_i^2$,  denoting $\partial _lq_i\equiv \dot q_i$, introducing the `mass tensor' ${\mathsf{m}}$ with positive $m_{1,2} $ and $m$, and parametrizing $y_i$ as below,
\begin{align*}
y_i&=e^{q_i}\,,\quad &{{\mathsf{m}}}&={\hat \kappa}^{-1}\equiv \left(
          \begin{array}{rr}
            m_1 & -m \\
            -m & m_2 \\
          \end{array}
        \right)\,,
\end{align*}
  we cast the second  of Eqs.~\eqref{Kondo}
into the  equations  of motion of {a} Newtonian particle  in a  2$D$ space,
\begin{align}\label{EoM}
{{\mathsf{m}}}\,\ddot{{\bm q}}&={\bm F}({\bm q})\,,
\end{align}
 with the initial conditions $  q_i(0){=}\tfrac{1}{2} \ln D_0 $ and $\dot{q}_i({0}){=-}\delta_i$.
The  corresponding   Lagrangian is
\begin{equation}\label{Lag}
L=\frac{1}{2}\,\dot{{\bm q}}\,{{\mathsf{m}}}\,\dot{{\bm q}}-U({\bm q})\,,\quad U({\bm q})=-\frac{1}{2}\sum_i\,e^{2q_i}\,.
\end{equation}
 Then  the total energy, $E=\frac{1}{2}\,\dot{{\bm q}}\,{{\mathsf{m}}}+U({{\bm{q}}}) $, is the integral of motion,
\begin{equation}
E=\frac{1}{2}\,\dot{{\bm q}}\,{{\mathsf{m}}}\,\dot{{\bm q}}-\frac{1}{2}\sum_i\,e^{2q_i}\,,
\end{equation}
given in terms of ${\bm{x}}$ and ${\bm{y}}$ in the main text, Eq.~\eqref{E0}.

If the mass tensor is diagonal, ${{\mathsf{m}}}=\mathrm{diag}[m_1\,,m_2]$, the Lagrangian would be a sum of two Lagrangians, $L=\sum_i\,L_i$, given by
\begin{equation}
L_i=\frac{1}{2}\,m_i\,\dot{q}_2^2+\frac{1}{2}\,e^{2q_i}.
\end{equation}
The explicit solution of the equations of motion,
\begin{equation}\label{solution}
D_i=e^{2q_i}=\frac{m_i\,k_i^2}{\sinh^2\left[k_i\,l+\chi_i\right]}\,,
\end{equation}
depends on {the} following parameters  found from the initial conditions:
\begin{equation*}
k_i^2=\delta_i^2-\frac{D_0}{m_i}\,,\quad \tanh\chi_i=\frac{k_i}{\delta_i}\,.
\end{equation*}
This solution has the  energy
$
E_i=\frac1{2}{m_i\,k_i^2}\,
$.
For positive energy the positive branch of the square root is assumed, $k_i=\sqrt{2E_i/m_i}$. As one can see, for the positive energy the disorder flows to zero (conducting phase):
\begin{equation}
D_i(l{\rightarrow}\infty) \rightarrow 0\,,\quad
E_i=\frac{1}{2}\left[\frac{\delta_i^2}{m_i}-D_0\right]>0\,.
\end{equation}
The same solution, Eq.~\eqref{solution}, is applicable for negative energies, $E_i\to-|E_i|$ and $k_i\to i\,|k_i|=i \sqrt{2|E_i|/m_i}$:
\begin{equation}\label{insulator}
D_i=e^{2q_i}=\frac{m_i\,|k_i|^2}{\sin^2\left[|k_i|\,l+\chi_i\right]}\,,
\end{equation}
with {the} necessary change $\chi_i\to i\chi_i$  to satisfy the initial conditions,
\begin{equation}
-|k_i|^2=\delta_i^2-\frac{D_0}{m_i}\,,\quad \tan\chi_i=\frac{|k_i|}{\delta_i}\,.
\end{equation}
This is {the} solution for insulating behavior that demonstrates divergence at a finite `{length}' $l_{\rm K}$, corresponding to the finite Kondo temperature.
Without coupling between channels, we have a 2$D$ Newtonian particle moving in a potential that is a sum of two parts, each depending {upon} one coordinate only. The degrees of freedom decouple and we have two uncoupled equations with two integrals of motion (energies for two directions), one for each equation.

\section{RG flows for  two coupled channels}
The coupling between channels introduces anisotropy in the dispersion law that couples motions in two perpendicular directions{, which} means that $E_1$ and $E_2$ are no longer   integrals of motion.  Now we have a single integral of motion, the total energy, Eq.~\eqref{E0}.
Nevertheless, for the weak coupling ($m\ll 1$) we may  {apply} adiabatic perturbation theory using the fact that both $E_i$'s will now be slowly changing:
\begin{equation}\label{adiabatic}
\dot{E}_i=m\,\dot{q}_i\,\ddot{q}_{-i}\,.
\end{equation}
Here we have used convention $-1\to2\,,-2\to1$ for the $-i$ index above. Since the energy is adiabatically changing, we can write approximate expressions for the solution, similar to the Eq.~\eqref{solution}:
 \begin{align}\label{solution_gen}
D_i(l)&=e^{2q_i(l)}=\frac{m_i\,k_i^2(l)}{\sinh^2\,\theta_i(l)}\,,\\
\theta_{i} (l)&=\int_0^{l}\,\mathrm{d}l'\,k_i(l')+\chi_i\,.\notag
\end{align}
 Which solution to use, conducting or insulating, depends on the energies $E_i=m_i\,k_i^2/2$. As they are not conserved due to the weak coupling,
we could start from the energy of one sign (e.g., positive, corresponding to the conducting behavior) but the correction could bring us into the negative territory where we have to switch to the insulating solution, as illustrated in Fig.~\ref{fig:CC}.

The adiabatic change of $E_{1,2} $ is governed by Eq.~\eqref{adiabatic},  which already contains {the} small parameter $m$, so that we can use {the uncoupled solutions from}  Eqs.~\eqref{solution} and \eqref{insulator} {there}. It is important to stress that $\ddot{q}_i\geq 0$ for both the solutions, $\dot{q}_i\leq 0$ for the conducting solution, while for the insulating solution it is positive if $\delta_i\leq 0$ and starts from negative for $\delta_i\geq 0$ but later turns positive.

These equations are not applicable in the strong coupling limit, where they give only a rough idea of the relevance of the Kondo physics: for  $\dot{q}_i>0$  {the} system flows towards {the} strong-coupling Kondo regime corresponding to an insulator. We therefore focus at the dynamics in the region with a starting point  $x_i\equiv \dot{q}_i<0$ corresponding to $\delta_i>0$ in Eq.~\eqref{RGxs}.

When both channels are initially in either conducting, $(cc)$,  or insulating, $(ii)$, phase, {a} rough estimate for the correction to the energies {gives}:
\begin{equation*}
\delta E_i(l)=\frac{m}{m_i}\left[D_i(l)-D_0\right]\,.
\end{equation*}
Then in the $(cc)$-phase, where $D_i(l\to\infty)\to0$,  both  initially positive energies acquire finite negative corrections,
\begin{equation}\label{guess}
\delta E_i=-\frac{m}{m_i}D_0=\frac{m_i}{2}\left[\delta_i^2- \left(\frac{1}{m_i}+\frac{m}{m_i^2}\right)D_0\right]\!,
\end{equation}
which simply shifts the boundary of stability further into the region of higher Luttinger parameters $K_{ii}$ so that the $(cc)$-phase shrinks. In the $(ii)$-phase we can still use Eq.~\eqref{guess} until both $D_i$ reach their minimum values (similar to the non-monotonic dependence of $J_{\perp}$ in the ferromagnetic Kondo system). Thus a negative energy acquires a negative correction so that there is no qualitative change:  the $(ii)$-phase remains itself.

In contrast to the two previous cases, the inter-channel interaction plays a crucial role for the initially mixed,  $({ic})$ phase, where one channel is conducting and one insulating.
In the region $\dot q<0$, we see that the positive energy of the conducting channel is suppressed by the correction while the negative energy of the insulating channel is boosted further into the negative territory. As before, the inter-channel coupling  {favors} insulating behavior and shrinks the $(ci)$-phase.

If we look for qualitative clues into the regime where $\dot q>0$  and disorder is growing, Eqs.~\eqref{insulator}, \eqref{solution_gen},   we see that when approaching the strong-coupling regime (the Kondo temperature) in the insulating channel, the correction to the conducting channel `energy' blows up. The conducting channel is thus blocked whilst the insulating one is strengthened, so that  the system moves towards the (ii)-phase analyzed above and found stable.

{To conclude, the mixed $(ci)$ phase is destroyed as the Kondo temperature is lowered whilst the (cc) and (ii) phase boundaries are shifted in favor of the insulator. Note that although inter-channel interactions force the Luttinger matrix to take higher values in order to arrive at a conducting state, the original Luttinger parameters can still be well below the threshold \(K>3/2\) in an isolated channel.}

%

\end{document}